\documentclass[12pt,letterpaper]{article}
\usepackage{amsmath,amssymb,array,calc,rotating,epsfig,psfrag,amscd, cite}

\makeatletter
\renewcommand\section{\@startsection {section}{1}{\z@}%
                                   {-3.5ex \@plus -1ex \@minus -.2ex}%nn
                                   {2.3ex \@plus.2ex}%
                                   {\normalfont\large\bfseries}}
\renewcommand\subsection{\@startsection{subsection}{2}{\z@}%
                                     {-3.25ex\@plus -1ex \@minus -.2ex}%
                                     {1.5ex \@plus .2ex}%
                                     {\normalfont\bfseries}}
\makeatother

\let\non\nonumber

\let\s=\sigma

\let\S=\Sigma

\newcommand{\bea}{\begin{eqnarray}}
\newcommand{\eea}{\end{eqnarray}}
\newcommand{\be}{\begin{equation}}
\newcommand{\ee}{\end{equation}}

\newcommand{\m}{\mu}
\newcommand{\n}{\nu}
\newcommand{\p}{\partial}

\newcommand{\C}[1]{$(\ref{#1})$}

\def\IZ{\relax\ifmmode\mathchoice
{\hbox{\cmss Z\kern-.4em Z}}{\hbox{\cmss Z\kern-.4em Z}}
{\lower.9pt\hbox{\cmsss Z\kern-.4em Z}} {\lower1.2pt\hbox{\cmsss
Z\kern-.4em Z}}\else{\cmss Z\kern-.4em Z}\fi}
\def\IR{\relax{\rm I\kern-.18em R}}

\def\one{{\hbox{ 1\kern-.8mm l}}}

\newlength{\bredde}
\def\slash#1{\settowidth{\bredde}{$#1$}\ifmmode\,\raisebox{.15ex}{/}
\hspace*{-\bredde} #1\else$\,\raisebox{.15ex}{/}\hspace*{-\bredde}
#1$\fi}

\newsavebox{\zzzbar}
\sbox{\zzzbar}
  {\setlength{\unitlength}{0.9em}
  \begin{picture}(0.6,0.7)
  \thinlines
  \put(0,0){\line(1,0){0.6}}
  \put(0,0.75){\line(1,0){0.575}}
  \multiput(0,0)(0.0125,0.025){30}{\rule{0.3pt}{0.3pt}}
  \multiput(0.2,0)(0.0125,0.025){30}{\rule{0.3pt}{0.3pt}}
  \put(0,0.75){\line(0,-1){0.15}}
  \put(0.015,0.75){\line(0,-1){0.1}}
  \put(0.03,0.75){\line(0,-1){0.075}}
  \put(0.045,0.75){\line(0,-1){0.05}}
  \put(0.05,0.75){\line(0,-1){0.025}}
  \put(0.6,0){\line(0,1){0.15}}
  \put(0.585,0){\line(0,1){0.1}}
  \put(0.57,0){\line(0,1){0.075}}
  \put(0.555,0){\line(0,1){0.05}}
  \put(0.55,0){\line(0,1){0.025}}
  \end{picture}}

\newcommand{\ena}{\end{eqnarray}}
\newcommand{\beqa}{\begin{eqnarray}}
\newcommand{\eeqa}{\end{eqnarray}}

\def\m{\mu}
\def\n{\nu}

\def\s{\sigma}

\def\S{\Sigma}

\begin{document}
\begin{titlepage}

\begin{center}

%\hfill \today
%\hfill         \phantom{xxx}         

%\hfill HRI

\vskip 2 cm
{\Large \bf A simplifying feature of the heterotic one loop four graviton amplitude }\\
\vskip 1.25 cm { Anirban Basu\footnote{email address:
    anirbanbasu@hri.res.in} } \\
{\vskip 0.5cm  Harish--Chandra Research Institute, HBNI, Chhatnag Road, Jhusi,\\
Allahabad 211019, India}

\end{center}

\vskip 2 cm

\begin{abstract}
\baselineskip=18pt

We show that the weight four modular graph functions that contribute to the integrand of the $t_8t_8D^4\mathcal{R}^4$ term at one loop in heterotic string theory do not require regularization, and hence the integrand is simple. This is unlike the graphs that contribute to the integrands of the other gravitational terms at this order in the low momentum expansion, and these integrands require regularization. This property persists for an infinite number of terms in the effective action, and their integrands do not require regularization. We find non--trivial relations between weight four graphs of distinct topologies that do not require regularization by performing trivial manipulations using auxiliary diagrams.

\end{abstract}

\end{titlepage}

%\pagestyle{plain}
%\baselineskip=18pt
% Try a wider skip
%\baselineskip=19pt
%%%%%%%%%%%%%%%%%%%%%%%%%%%%%%%%%%%%%%%%%%%%%%%%%%%%%%%%%%%%%%%%%%%%%%%%%%%%%%

The expression for the four graviton amplitude in type II string theory takes a very compact and simple form. Its low momentum expansion leads to terms in the effective action that are analytic and non--analytic in the external momenta of the on--shell particles. The tensor structure of the analytic contribution is schematically denoted as $t_8 t_8 D^{2k} \mathcal{R}^4$ at a fixed order in the $\alpha'$ expansion, which arises from the structure of the scattering amplitudes at tree level, and at one, two and three loops. Here $D$ schematically refers to a derivative, while $\mathcal{R}$ refers to the Riemann tensor. Thus at every order in the $\alpha'$ expansion, from known results in string perturbation theory upto three loops~\cite{Green:1981xx,Green:1981ya,D'Hoker:2005vch,Gomez:2013sla}, we see that the purely gravitational terms are the bosonic components of only one superinvariant in the IIA/IIB theory\footnote{The perturbative part of the four graviton amplitude is the same in the IIA and IIB theories upto two loops~\cite{Green:1999pu}, and upto the $D^8\mathcal{R}^4$ term at higher loops~\cite{Berkovits:2006vc,Gomez:2013sla}. Beyond that they differ for the IIA and IIB theories at higher loops as the $\epsilon_{10}\epsilon_{10} D^{2k}\mathcal{R}^4$ terms contribute coming from the odd--odd spin structures. This difference shall not concern us.}.   

This simplicity is related to the enormous supersymmetry the type II theory enjoys in ten dimensions or in toroidal compactifications without breaking supersymmetry. Theories with less supersymmetry lead to terms in the effective action such that the purely gravitational terms are the bosonic components of several superinvariants, the structure of which is determined by the details of the theory. The heterotic string theory~\cite{Gross:1984dd} in ten dimensions which has half maximal supersymmetry is one of the simplest settings where these issues can be studied in detail. Considering terms in the effective action that follow from the low momentum expansion of the four graviton amplitude, these superivariants have been studied for the eight derivative interactions~\cite{Bergshoeff:1989de,deRoo:1992zp}, and this structure is expected to generalize to higher derivative interactions in the effective action. Now among these superinvariants, we have the one whose bosonic component is the $t_8 t_8 D^{2k} \mathcal{R}^4$ term and this is the only purely gravitational term whose tensor structure is the same in the type II and the heterotic theories. 

The coefficients of the various tensors, and hence their contributions to the effective action, are uniquely determined by the string amplitude. At one loop, this is given by a modular invariant integral over the truncated fundamental domain of $SL(2,\mathbb{Z})$. In the type II theory, the integrand involves $SL(2,\mathbb{Z})$ invariant modular graph functions. The vertices of these graphs are given by the positions of insertions of vertex operators on the toroidal worldsheet, and the links are given by scalar Green functions. Various equations have been derived for these graphs which are needed to calculate their contribution to the type II amplitudes~\cite{Green:1999pv,Green:2008uj,D'Hoker:2015foa,D'Hoker:2015zfa,Basu:2015ayg,Zerbini:2015rss,DHoker:2015wxz,Basu:2016fpd,D'Hoker:2016jac,Basu:2016xrt,Basu:2016kli,Basu:2016mmk,DHoker:2016quv,Kleinschmidt:2017ege,DHoker:2017zhq}, based on Poisson equations they satisfy and asymptotic expansions. In the heterotic theory, the integrand involves $SL(2,\mathbb{Z})$ covariant graphs instead, contributing in modular invariant combinations to the integrand that are determined by the details of the structure of the amplitude~\cite{Ellis:1987dc,Abe:1988cq,Basu:2017nhs}. Some or all of the links in several of these graphs are given by the derivative of the Green function. Hence the structure of the integrand is richer than that in the type II theory. Now since the $t_8 t_8 D^{2k} \mathcal{R}^4$ term is common to both the type II and the heterotic theories, it is natural to expect that the integrand of this term might be simpler than the integrands of the other gravitational terms in the heterotic theory that arise from other superinvariants. If this is the case, we would also like to know in what sense this integrand is simpler. We analyze these issues for the case of the $t_8 t_8 D^4 \mathcal{R}^4$ term in the effective action of the heterotic theory, and indeed find that the integrand is simpler than the ones for the other gravitational terms at the same order in the derivative expansion. The graphs of modular weight four that arise in the integrand for this interaction do not involve closed loops of a certain kind, and hence do not need regularization in defining them, leading to the simplicity. The graphs that arise for the other interactions all do not have this property. We also discuss when such simplification occurs at higher orders in the momentum expansion. 

It will be interesting to have a general understanding of the basis elements of modular covariant graphs. We also expect richer structures to emerge at higher loops, which will be worth unravelling.   

In our analysis, we shall consider the four graviton amplitude at tree level and at one loop in the heterotic theory, which is the same in the ${\rm Spin}(32)/\mathbb{Z}_2$ and the $E_8\times E_8$ theories.
For this amplitude, $\epsilon^{(i)}_{\mu\nu}$ ($i=1,\cdots, 4$) is the polarization tensor for the graviton which carries momentum $k_i$, and $\phi$ is the dilaton. The Mandelstam variables are given by
\be s= -(k_1+k_2)^2, \quad t= -(k_1 + k_4)^2, \quad u = -(k_1+k_3)^2,\ee
which satisfy $s+t+u=0$. Explicitly, the $t_8t_8\mathcal{R}^4$ tensor is given by
\be \label{tensor}t_8 t_8 \mathcal{R}^4 \equiv \prod_{i=1}^4\epsilon^{(i)}_{\mu_i\nu_i} K^{\mu_1 \mu_2 \mu_3 \mu_4}  K^{\nu_1\nu_2\nu_3\nu_4},\ee
where the crossing symmetric tensor $K^{\m_1\m_2\m_3\m_4}$ is given by
\bea \label{defK}K^{\m_1\m_2\m_3\m_4}=\frac{1}{4} (ut \eta^{\m_1\m_2} \eta^{\m_3\m_4} + st\eta^{\m_1\m_3}\eta^{\m_2\m_4} + su \eta^{\m_1\m_4}\eta^{\m_2\m_3})\non \\ -\frac{s}{2} (\eta^{\m_2\m_4} k_4^{\m_1}k_2^{\m_3} +\eta^{\m_1\m_3} k_1^{\m_4} k_3^{\m_2} +\eta^{\m_2\m_3} k_2^{\m_4} k_3^{\m_1} + \eta^{\m_1\m_4} k_4^{\m_2} k_1^{\m_3})\non \\ -\frac{t}{2} (\eta^{\m_2\m_4} k_2^{\m_1}k_4^{\m_3} +\eta^{\m_1\m_3} k_3^{\m_4} k_1^{\m_2} +\eta^{\m_1\m_2} k_2^{\m_4} k_1^{\m_3} + \eta^{\m_3\m_4} k_3^{\m_1} k_4^{\m_2})\non \\ -\frac{u}{2} (\eta^{\m_1\m_2} k_1^{\m_4}k_2^{\m_3} +\eta^{\m_2\m_3} k_3^{\m_4} k_2^{\m_1} +\eta^{\m_1\m_4} k_1^{\m_2} k_4^{\m_3} + \eta^{\m_3\m_4} k_3^{\m_2} k_4^{\m_1}).\eea

First we briefly consider the four graviton amplitude in the heterotic theory at tree level. It is given by\cite{Gross:1985rr,Kawai:1985xq,Cai:1986sa,Gross:1986mw,Kikuchi:1986cz}
\bea \label{t1}&&A^{tree}_{Het} (k_i, \epsilon^{(i)})= -e^{-2\phi}\frac{\Gamma(-\alpha' s/4)\Gamma(-\alpha' t/4)\Gamma(-\alpha' u/4)}{\Gamma(1+\alpha' s/4)\Gamma(1+\alpha' t/4)\Gamma(1+\alpha' u/4)}\prod_{i=1}^4 \epsilon^{(i)}_{\mu_i\nu_i} K^{\mu_1 \mu_2 \mu_3 \mu_4} \non \\ &&\times \Big[ K^{\nu_1\nu_2\nu_3\nu_4} -\frac{\alpha'stu}{16}\Big( \frac{\eta^{\nu_1\nu_2} \eta^{\nu_3\nu_4}}{1+\alpha's/4}+ \frac{\eta^{\nu_1\nu_4} \eta^{\nu_2\nu_3}}{1+\alpha't/4}+ \frac{\eta^{\nu_1\nu_3} \eta^{\nu_2\nu_4}}{1+\alpha'u/4}\Big)\non \\ &&-\frac{\alpha' ut}{8(1+\alpha's/4)}\Big( \eta^{\n_1\nu_2} k_3^{\n_4} k_4^{\n_3} + \eta^{\n_3\n_4} k_1^{\n_2} k_2^{\n_1}\Big) -\frac{\alpha' us}{8(1+\alpha't/4)}\Big( \eta^{\n_1\nu_4} k_2^{\n_3} k_3^{\n_2}+ \eta^{\n_2\n_3} k_1^{\n_4} k_4^{\n_1}\Big)\non \\ &&-\frac{\alpha' st}{8(1+\alpha'u/4)}\Big(\eta^{\n_1\nu_3} k_2^{\n_4} k_4^{\n_2}+ \eta^{\n_2\n_4} k_1^{\n_3} k_3^{\n_1}\Big)  +\ldots\Big],\eea
where we have ignored terms having four factors of $k_i^{\n_j}$ only (for example, $\alpha's k_1^{\n_3} k_2^{\n_4} k_4^{\n_1} k_3^{\n_2}/12$) as they are not relevant for our purposes. The first term in \C{t1} gives the $t_8t_8\mathcal{R}^4$ tensor while the remaining terms yield bosonic contributions to other supermultiplets\footnote{We keep these terms at tree level as they are useful in organizing the multiplets at one loop. Unlike at tree level, where the $t_8t_8 D^{2k}\mathcal{R}^4$ terms are easily obtained before performing the low momentum expansion, at one loop we obtain this separation only after performing the low momentum expansion. }.

The low momentum expansion of \C{t1} yields contact interactions with at least eight powers of momenta. Keeping only these terms, the $\mathcal{R}^4$  interaction in the effective action is obtained from
\bea \label{tree1}A^{tree}_{Het,\mathcal{R}^4} = e^{-2\phi} \prod_{i=1}^4 \epsilon^{(i)}_{\mu_i\nu_i} K^{\mu_1 \mu_2 \mu_3 \mu_4}   \Big[ 2\zeta(3)K  -L_1\Big]^{\n_1\n_2\n_3\n_4}, \eea
while the $D^2 \mathcal{R}^4$ and $D^4\mathcal{R}^4$ terms are obtained from
\bea \label{tree2}A^{tree}_{Het,D^2\mathcal{R}^4} = e^{-2\phi} \prod_{i=1}^4 \epsilon^{(i)}_{\mu_i\nu_i} K^{\mu_1 \mu_2 \mu_3 \mu_4} \Big[ L_2  -2\zeta(3) M_0\Big]^{\n_1\n_2\n_3\n_4}  \eea
and 
\bea \label{tree3}A^{tree}_{Het,D^4\mathcal{R}^4} = e^{-2\phi} \prod_{i=1}^4 \epsilon^{(i)}_{\mu_i\nu_i} K^{\mu_1 \mu_2 \mu_3 \mu_4} \Big[ \zeta(5)\s_2  K -L_3   +2\zeta(3) M_1 \Big]^{\n_1\n_2\n_3\n_4}. \eea
respectively, where we have defined
\be \s_k =  \Big(\frac{\alpha'}{4}\Big)^k (s^k + t^k + u^k).\ee 
Thus from \C{tree1}, \C{tree2} and \C{tree3} we see that the $t_8t_8D^{2k} \mathcal{R}^4$ contributions are given by
\be A^{tree}_{Het} = e^{-2\phi} t_8t_8 \mathcal{R}^4 \Big[ 2\zeta(3) +\zeta(5)\s_2\Big]\ee
upto this order in the derivative expansion\footnote{The $t_8t_8D^2\mathcal{R}^4$ contribution vanishes kinematically using $\s_1=0$.}. The remaining terms in these expressions arise from different supermultiplets which are given, in general, by  the tensors\footnote{These also have contributions involving four powers of $k_i^{\n_j}$ which we have ignored. These are explicitly given for the $\mathcal{R}^4$ and $D^2\mathcal{R}^4$ terms at tree level and at one loop in~\cite{Ellis:1987dc,Basu:2017nhs}.}          
\bea \label{T1}L_n^{\m_1\m_2\m_3\m_4} &=& \frac{2}{\alpha'}\Big(I_{2;n,0} +\frac{2}{\alpha'} I_{1;n+1,0}\Big)^{\m_1\m_2\m_3\m_4}, \non \\ M_n^{\m_1\m_2\m_3\m_4} &=& \frac{2}{\alpha'}\Big(I_{2;n,1} + \frac{2\s_3}{3\alpha'} I_{1;n,0}\Big)^{\m_1\m_2\m_3\m_4}.\eea
We also define the tensor
\bea \label{T2}K_n^{\m_1\m_2\m_3\m_4} &=& \Big(I_{3,n} - \frac{8}{\alpha'^2} I_{1;n,1}\Big)^{\m_1\m_2\m_3\m_4} \eea
which will be useful later. In \C{T1} and \C{T2}, the various crossing symmetric tensors are given by  
\bea I_{1;m,n}^{\m_1\m_2\m_3\m_4} &=& \Big(\frac{\alpha's}{4}\Big)^m \Big(\frac{\alpha'^2 ut}{16}\Big)^n \eta^{\m_1\m_2} \eta^{\m_3\m_4} + \Big(\frac{\alpha't}{4}\Big)^m \Big(\frac{\alpha'^2 su}{16}\Big)^n \eta^{\m_1\m_4}\eta^{\m_2\m_3} \non \\ &&+\Big(\frac{\alpha'u}{4}\Big)^m \Big(\frac{\alpha'^2 st}{16}\Big)^n\eta^{\m_1\m_3}\eta^{\m_2\m_4},\eea
which has two factors of the metric, and
\bea  I_{2;m,n}^{\m_1\m_2\m_3\m_4} &=& \Big(\frac{\alpha's}{4}\Big)^m \Big(\frac{\alpha'^2 ut}{16}\Big)^n (\eta^{\m_1\m_2} k_3^{\m_4} k_4^{\m_3} + \eta^{\m_3\m_4} k_1^{\m_2} k_2^{\m_1})  \non \\ && +  \Big(\frac{\alpha't}{4}\Big)^m \Big(\frac{\alpha'^2 su}{16}\Big)^n(\eta^{\m_1\m_4} k_2^{\m_3} k_3^{\m_2} + \eta^{\m_2\m_3} k_1^{\m_4} k_4^{\m_1})\non \\ &&+   \Big(\frac{\alpha'u}{4}\Big)^m \Big(\frac{\alpha'^2 st}{16}\Big)^n(\eta^{\m_1\m_3} k_2^{\m_4} k_4^{\m_2} + \eta^{\m_2\m_4} k_1^{\m_3} k_3^{\m_1}), \non \\ I_{3,n}^{\m_1\m_2\m_3\m_4} &=& \Big(\frac{\alpha's}{4}\Big)^n  \Big[\eta^{\m_1\m_2}(tk_1^{\m_3} k_2^{\m_4} + u k_1^{\m_4} k_2^{\m_3}) + \eta^{\m_3\m_4} (t k_3^{\m_1} k_4^{\m_2} + u k_3^{\m_3} k_4^{\m_1})\Big]\non \\ && +\Big(\frac{\alpha't}{4}\Big)^n  \Big[\eta^{\m_1\m_4}(sk_1^{\m_3} k_4^{\m_2} + u k_1^{\m_2} k_4^{\m_3}) + \eta^{\m_2\m_3} (s k_2^{\m_4} k_3^{\m_1} + u k_2^{\m_1} k_3^{\m_4})\Big]\non \\ && +\Big(\frac{\alpha'u}{4}\Big)^n  \Big[\eta^{\m_1\m_3}(sk_1^{\m_4} k_3^{\m_2} + t k_1^{\m_2} k_3^{\m_4}) + \eta^{\m_2\m_4} (s k_2^{\m_3} k_4^{\m_1} + t k_2^{\m_1} k_4^{\m_3})\Big] \eea 
which have one factor of the metric. 
Thus we see that
\be  K^{\m_1\m_2\m_3\m_4} = -\frac{1}{2} K_0^{\m_1\m_2\m_3\m_4} .\ee
This simple analysis of the decomposition of the expression for the tree level amplitude into various supermultiplets generalizes at one loop as we now see below. 

The one loop four graviton amplitude is given by~\cite{Gross:1985rr,Sakai:1986bi,Ellis:1987dc,Abe:1988cq}
\bea \label{1loop4g}A^{1-loop}_{Het} (k_i, \epsilon^{(i)}) = \prod_{i=1}^4 \epsilon^{(i)}_{\mu_i\nu_i} K^{\mu_1 \mu_2 \mu_3 \mu_4} \int_{\mathcal{F}} \frac{d^2\tau}{\tau_2^2} \frac{{\bar{E}}_4^2}{{\bar{\eta}}^{24}}\prod_{i=1}^4 \int_\S  \frac{d^2 z^i}{\tau_2} e^{\mathcal{D}} \mathcal{T}^{\n_1\n_2\n_3\n_4}, \eea
where we have defined
\be 2\zeta(2k) E_{2k}(\tau) = G_{2k} (\tau)\ee
for $k \geq 2$, where $G_{2k}(\tau)$ is the holomorphic Eisenstein series of modular weight $2k$ defined by
\be \label{defG}G_{2k} (\tau) = \sum_{(m,n)\neq (0,0)} \frac{1}{(m+n\tau)^{2k}}.\ee
In \C{1loop4g} we have integrated over $\mathcal{F}$, the fundamental domain of $SL(2,\mathbb{Z})$ and $d^2 \tau = d\tau_1 d\tau_2$.  The insertion points of the vertex operators $z_i$ $(i=1,\cdots,4)$ on the toroidal worldsheet $\S$  with complex structure $\tau$ are integrated over. We have that $d^2 z_i = d({\rm Re} z_i) d({\rm Im}z_i)$, where
\be -\frac{1}{2} \leq {\rm Re} z_i \leq \frac{1}{2}, \quad 0 \leq {\rm Im} z_i\leq \tau_2 \ee
for all $i$. The Koba--Nielsen factor, $\mathcal{D}$ is defined by
\be 4 \alpha'^{-1}\mathcal{D} = s (G_{12} + G_{34}) + t (G_{14} + G_{23}) +u (G_{13} +G_{24}) \ee
where ${G}_{ij}$ is the scalar Green function between points $z_i$ and $z_j$, and hence
\be G_{ij} \equiv {G}(z_i - z_j;\tau).\ee
It is explicitly given by~\cite{Lerche:1987qk,Green:1999pv}
\bea \label{Green}G(z;\tau) =\frac{1}{\pi} \sum_{(m,n)\neq(0,0)} \frac{\tau_2}{\vert m\tau+n\vert^2} e^{\pi[\bar{z}(m\tau+n)-z(m\bar\tau+n)]/\tau_2} .\eea
Finally, the crossing symmetric tensor $\mathcal{T}^{\m_1\m_2\m_3\m_4}$ is defined by
\bea \label{T}&&\mathcal{T}^{\m_1\m_2\m_3\m_4} = A^{\m_1} A^{\m_2} A^{\m_3} A^{\m_4} +\frac{1}{2\alpha'} \Big(\eta^{\m_1\m_2} R_{12} A^{\m_3} A^{\m_4} + \eta^{\m_1\m_3} R_{13} A^{\m_2} A^{\m_4} \non \\ &&+\eta^{\m_1\m_4} R_{14} A^{\m_2}A^{\m_3} + \eta^{\m_2\m_3} R_{23} A^{\m_1} A^{\m_4} +\eta^{\m_2\m_4} R_{24} A^{\m_1} A^{\m_3} +\eta^{\m_3\m_4} R_{34} A^{\m_1}A^{\m_2}\Big)\non \\ &&+\frac{1}{(2\alpha')^2} \Big(\eta^{\m_1\m_2}\eta^{\m_3\m_4}R_{12}R_{34} + \eta^{\m_1\m_3}\eta^{\m_2\m_4}R_{13}R_{24} +\eta^{\m_1\m_4}\eta^{\m_2\m_3}R_{14}R_{23}\Big),\eea
where
\bea \label{defAR}A^{\mu_i} &=& \frac{1}{4\pi i}\sum_{j=1}^4 k_j^{\m_i} \bar\p_j G_{ji}, \non \\ R_{ij} &=& -\frac{1}{4\pi^2} \bar\p^2_i G_{ij},\eea
which involve derivatives of the Green functions.

Now the analytic part of the amplitude \C{1loop4g} is obtained by expanding the Koba--Nielsen factor in powers of $\alpha'$ and integrating over the truncated fundamental domain $\mathcal{F}_L$ defined by~\cite{Green:1999pv,Green:2008uj}
\be \mathcal{F}_L = \{ -\frac{1}{2} \leq \tau_1 \leq \frac{1}{2}, \quad \vert \tau \vert \geq 1, \quad \tau_2 \leq L\}\ee
and keeping the finite terms as $L\rightarrow \infty$.

We now perform the low momentum expansion of the amplitude and express the integrand in terms of modular graph functions. The single--valuedness of the Green function allows us to integrate $\p_z G(z,w)$ by parts while integrating over $z$ without picking up boundary terms. From \C{Green} it also follows that all one particle reducible graphs vanish, and
\be \int_\S d^2 z \p_z G(z,w) = \int_\S d^2 z \p_z^2 G(z,w)=0.\ee    
Also there are no graphs where a vertex has only one link ending on it. 

Finally, the Green function equations are
\bea \label{eigen}\bar{\p}_w\p_z G(z,w) = \pi \delta^2 (z-w) - \frac{\pi}{\tau_2}, \non \\ \bar{\p}_z\p_z G(z,w) = -\pi \delta^2 (z-w) + \frac{\pi}{\tau_2} \eea
which we use frequently.

In the various graphs below, while the black links are simply the Green functions, the conventions for the others involving derivatives of Green functions is given in figure 1. 
\begin{figure}[ht]
\begin{center}
\[
\mbox{\begin{picture}(280,60)(0,0)
\includegraphics[scale=.75]{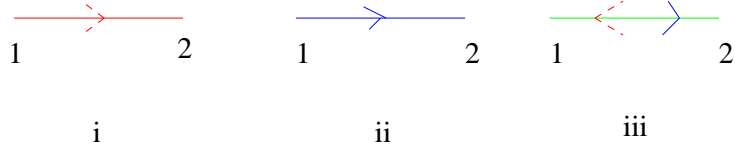}
\end{picture}}
\]
\caption{ (i) $\p_2 G_{12} = -\p_1 G_{12}$, (ii) $\bar\p_2 G_{12} = -\bar\p_1 G_{12}$ and (iii) $\p_1 \bar\p_2 G_{12}$}
\end{center}
\end{figure}

In performing the low momentum expansion from the expression \C{T} there are contributions of the form $A^4$, $A^2 R/\alpha'$ and $R^2/\alpha'^2$ schematically from $\mathcal{T}$. Each of them must be multiplied by a term involving appropriate powers of the external momenta that result from expanding the Koba--Nielsen factor $e^{\mathcal{D}}$ as a polynomial in $\mathcal{D}$. We ignore the contributions coming from $A^4$ as they are not relevant for our purposes. Now the first non--vanishing term in the low momentum expansion of \C{1loop4g} yields the $\mathcal{R}^4$ term.    

For the various terms, we denote
\be \label{1G}A^{1-loop}_{D^{2k}\mathcal{R}^4} (k_i, \epsilon^{(i)}) = \prod_{i=1}^4 \epsilon^{(i)}_{\mu_i\nu_i} K^{\mu_1 \mu_2 \mu_3 \mu_4} \int_{\mathcal{F}_L} \frac{d^2\tau}{\tau_2^2} \frac{{\bar{E}}_4^2}{{\bar{\eta}}^{24}} \mathcal{X}^{\n_1\n_2\n_3\n_4}_{D^{2k}\mathcal{R}^4}\ee
where $\mathcal{X}^{\n_1\n_2\n_3\n_4}_{D^{2k}\mathcal{R}^4}$ is given below for the various interactions, and we have integrated over the truncated fundamental domain $\mathcal{F}_L$. 

For the $\mathcal{R}^4$ term, we have that~\cite{Ellis:1987dc,Abe:1988cq}
\bea \mathcal{X}^{\n_1\n_2\n_3\n_4}_{\mathcal{R}^4} = -\frac{1}{(4\pi)^4} \Big[ 2 Q_1 K - (Q_1 + Q_2^2) L_1 \Big]^{\nu_1\nu_2\nu_3\nu_4}.\eea

\begin{figure}[ht]
\begin{center}
\[
\mbox{\begin{picture}(160,100)(0,0)
\includegraphics[scale=.65]{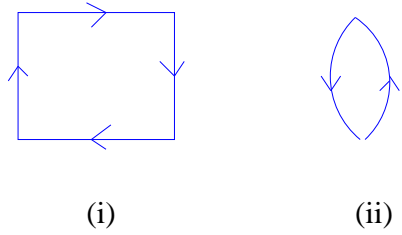}
\end{picture}}
\]
\caption{Graphs for $\mathcal{R}^4$: (i) $Q_1$ and (ii) $Q_2$}
\end{center}
\end{figure}
The relevant graphs are given in figure 2~\footnote{The vertices of all graphs in our analysis are integrated with the measure $\int_{\S} d^2z/\tau_2$.}. Thus for the $\mathcal{R}^4$ term, we see that there are two supermultiplets which contribute, one being $t_8t_8\mathcal{R}^4$. Though the graphs that enter the analysis are not complicated, there is an important distinction between $Q_1$ and $Q_2$. The graph $Q_2$ involves $(\bar\p_i G_{ij})^2$ where $i,j$ are integrated over. Graphs involving such factors are naively divergent and have to be defined by regularization. In fact, we have that
\be 2\zeta(2) Q_2 = {\rm lim}_{s\rightarrow 0} \sum_{(m,n) \neq (0,0)} \frac{1}{(m+n\tau)^2\vert m+n\tau\vert^{2s}} = \overline{G_2(\tau)} - \frac{\pi}{\tau_2}\ee
on using \C{defG}.     
Thus the regularization introduces non--holomorphicity while producing a weight 2 modular form. $Q_1$ on the other hand does not require any regularization, and is given by $2\zeta(4) \overline{E_4(\tau)}$. 

Thus for the $\mathcal{R}^4$ contribution, we see that the $t_8t_8\mathcal{R}^4$ contribution is simple as it does not involve any graphs which require regularization, which is not so for the other term involving $L_1$. Of course, here the graph $Q_2$ is elementary, and at higher orders in the momentum expansion, we shall encounter complicated graphs. However, what remains true is that graphs having factors of $(\bar\p_i G_{ij})^2$ require regularization. This can be easily seen by performing an asymptotic expansion in large $\tau_2$ as was done for some cases in~\cite{Basu:2017nhs}, and looking at the contributions that are power behaved in $\tau_2$ by setting appropriate lattice momenta to zero and performing the sum by Poisson resummation. Of course, calculating the coefficients of the power behaved terms in $\tau_2$ in the asymptotic expansion need not be easy, but the fact that regularization is needed is easy to see. 

Does this simplicity for the $t_8t_8\mathcal{R}^4$ term extend to the graphs at higher orders in the momentum expansion? This is the question we would like to address in detail for the $D^4\mathcal{R}^4$ term in the momentum expansion.           
  
Hence, we next consider the $D^2\mathcal{R}^4$ term, where we have that~\cite{Basu:2017nhs}
\bea &&\mathcal{X}^{\n_1\n_2\n_3\n_4}_{D^2\mathcal{R}^4} = \frac{1}{(4\pi)^4} \Big[2 (Q_2 Q_3 - Q_5) L_2 +2(Q_2 Q_4 + Q_5 + 2 Q_6 - 3 Q_7 - 2 Q_8) M_0 \non \\ &&-(Q_2 Q_4 + Q_5 - Q_6 + 3 Q_7 + Q_8)K_1\Big]^{\n_1\n_2\n_3\n_4}. \eea
The relevant graphs are given in figure 3. 
\begin{figure}[ht]
\begin{center}
\[
\mbox{\begin{picture}(240,200)(0,0)
\includegraphics[scale=.8]{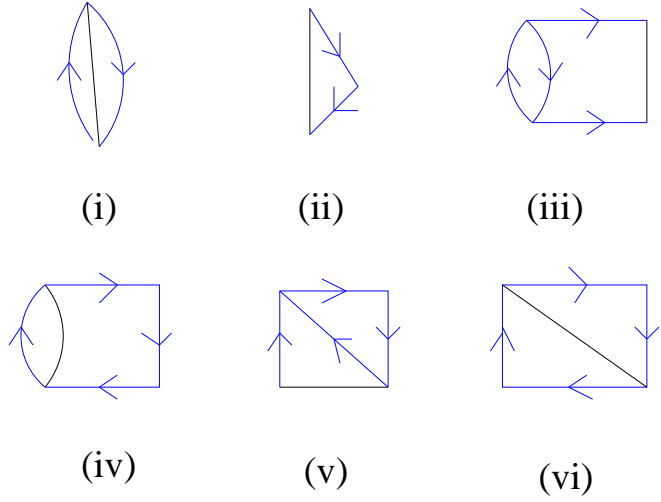}
\end{picture}}
\]
\caption{New graphs for $D^2\mathcal{R}^4$: (i) $Q_3$, (ii) $Q_4$, (iii) $Q_5$, (iv) $Q_6$, (v) $Q_7$ and (vi) $Q_8$}
\end{center}
\end{figure}
While the $t_8t_8D^2\mathcal{R}^4$ term vanishes for kinematical reasons, the other superinvariants involving $L_2$, $M_0$ and $K_1$ involve graphs that require regularization. Note that the tensor involving $K_1$ is absent at tree level in \C{tree2}.

We finally consider the $D^4\mathcal{R}^4$ term in detail. From the $A^2 R e^{\mathcal{D}}/\alpha'$ term in the integrand we get
\bea \frac{1}{3\alpha'(4\pi)^4} \Big[ P_1 I_{2;3,0} + P_2 I_{2,1;1} + \frac{\alpha'P_3 \s_2}{8} I_{3,0} + \frac{\alpha'}{4}(P_4 - P_3)I_{3,2}\Big]^{\n_1\n_2\n_3\n_4},\eea 
while the $R^2 e^{\mathcal{D}}/\alpha'^2$ term gives 
\bea  \frac{1}{3(4\pi)^4\alpha'^2} \Big[ 2 P_1 I_{1;4,0} + \frac{2P_2\s_3}{3}I_{1,1;0} -P_3 \s_2 I_{1;0,1} - 2(P_4 - P_3)I_{1;2,1}\Big]^{\n_1\n_2\n_3\n_4},\eea
where $P_i$ involves the various graphs given by
\bea \label{Pi}P_1 &=& 12 Q_{11} + 12 Q_{12} + 6 Q_2 Q_9 + 6 Q_3^2 + 6 \mathcal{E}_2 Q_2^2 - 12 Q_{13} + 6 Q_{14} - 6 Q_{15}, \non  \\ P_2 &=& 6 Q_{16} -12 Q_{11} - 18 Q_{12} -12 \mathcal{E}_2 Q_2^2 + 12 Q_3 Q_4 +24 Q_2 Q_{10} +6Q_{17} \non \\ &&-6 Q_{18} +30 Q_{13}-12 Q_{14} + 24 Q_{15} - 24 Q_{19} -36 Q_{20} + 24 Q_{21}, \non \\ P_3 &=& -48 Q_{19} +12Q_{17} +12 Q_{12} +12Q_{13} +12Q_{16}, \non  \\ P_4 &=& -12 Q_3 Q_4 + 6 Q_{16} + 18 Q_{12} +12Q_{11} + 6 Q_{17} -6 Q_{18} - 6 Q_{13} \non \\ && -24 Q_{19} + 24 Q_{21} - 12 Q_{20}.\eea
In \C{Pi}, the graph $\mathcal{E}_2$ is given in figure 4, while the others are given in figure 5. 

\begin{figure}[ht]
\begin{center}
\[
\mbox{\begin{picture}(40,80)(0,0)
\includegraphics[scale=.7]{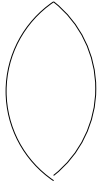}
\end{picture}}
\]
\caption{Graph for $\mathcal{E}_2$}
\end{center}
\end{figure}

In obtaining \C{Pi}, we have also used the relations between the graphs given in figure 6~\cite{Basu:2017nhs}. 
Thus adding the various contributions we get that 
\bea \label{D4R4}\mathcal{X}^{\n_1\n_2\n_3\n_4}_{D^4\mathcal{R}^4} = \frac{1}{3!(4\pi)^4} \Big[ P_1 L_3 + P_2 M_1 - \frac{P_3\s_2}{2} K+ \frac{P_4 - P_3}{2}K_2\Big]^{\n_1\n_2\n_3\n_4} . \eea

\begin{figure}[ht]
\begin{center}
\[
\mbox{\begin{picture}(340,320)(0,0)
\includegraphics[scale=.7]{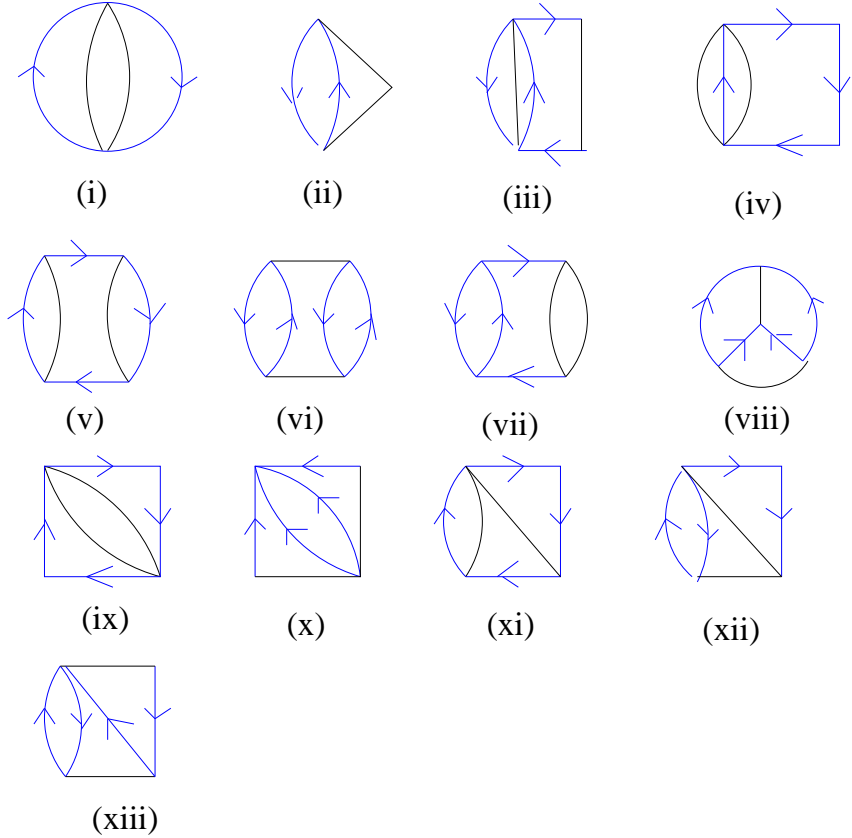}
\end{picture}}
\]
\caption{New graphs for $D^4\mathcal{R}^4$: (i) $Q_9$, (ii) $Q_{10}$, (iii) $Q_{11}$, (iv) $Q_{12}$, (v) $Q_{13}$, (vi) $Q_{14}$, (vii) $Q_{15}$, (viii) $Q_{16}$, (ix) $Q_{17}$, (x) $Q_{18}$, (xi) $Q_{19}$, (xii) $Q_{20}$ and (xiii) $Q_{21}$}
\end{center}
\end{figure}

Note that the tensor involving $K_2$ in \C{D4R4} is absent in \C{tree3}. Also, the $t_8t_8 D^4\mathcal{R}^4$ term involves the combination $P_3$, which has graphs which do not need regularization, and hence is simple. On the other hand, the other supermultiplets involve the combinations $P_1$, $P_2$ and $P_4$ all of which involve graphs that require regularization.

\begin{figure}[ht]
\begin{center}
\[
\mbox{\begin{picture}(190,150)(0,0)
\includegraphics[scale=.7]{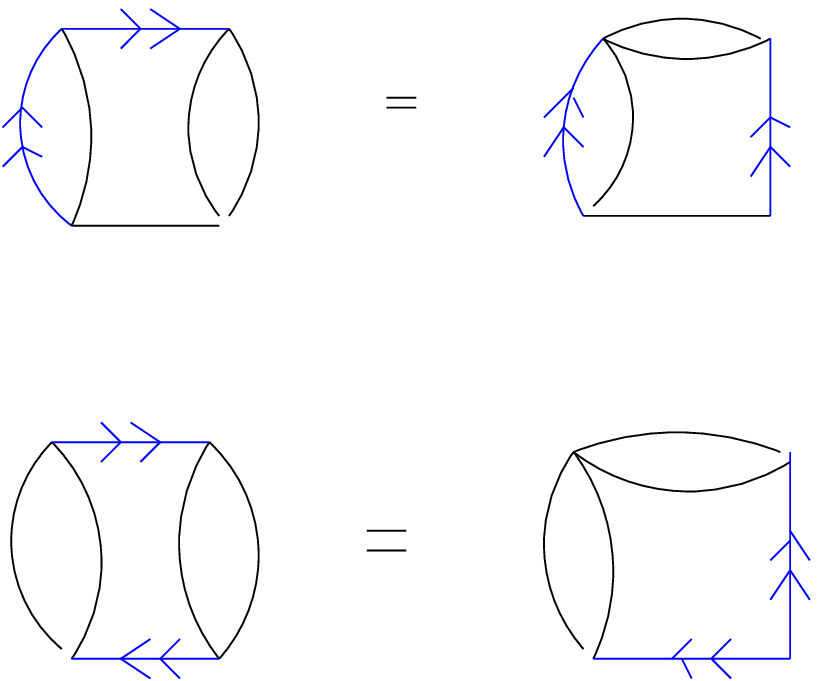}
\end{picture}}
\]
\caption{Some relations between graphs}
\end{center}
\end{figure}

Thus we see that the $t_8t_8\mathcal{R}^4$ and $t_8t_8D^4\mathcal{R}^4 = t_8t_8 \s_2 \mathcal{R}^4$ terms have integrands which are simple, in the sense that they only involve graphs that do not require regularization. We have not done a detailed analysis of the supermultiplet structure at higher orders in the momentum expansion. However, the kinematic structure of the various terms in the amplitude gives us some results at arbitrary orders in the momentum expansion. By eliminating $u$ and considering graphs which contribute to the $t^{2l+2} \eta^{\n_1\n_2}\eta^{\n_3\n_4}$ term\footnote{Apart from the factor of $\prod_{i=1}^4 \epsilon^{(i)}_{\m_i\n_i} K^{\m_1\m_2\m_3\m_4}$ which is always present.}, we see that the integrands of the $t_8t_8\s_2^{l} \mathcal{R}^4$ terms are simple. Also by considering graphs which contribute to the $s^m t^{2(l+m+1)}\eta^{\n_1\n_2}\eta^{\n_3\n_4}$ ($m\neq 0$) term on eliminating $u$, we see that the integrands of the $t_8t_8\s_2^l \s_3^m\mathcal{R}^4$ terms for $m\neq 0$ need not be simple\footnote{These two cases exhaust all the $t_8t_8D^{2k} \mathcal{R}^4$ terms using $\s_1=0$.}.    

Thus we see that an infinite class of terms in the effective action have simple integrands, which raises an interesting question which we now address. First let us consider the low momentum expansion of the tree level amplitudes. In various cases where this expansion has been performed, the coefficients of the various terms are given by multi--zeta values whose transcendentality increases as one considers terms at higher orders in the momentum expansion (see~\cite{Schlotterer:2012ny}, for example). In order to simplify calculations and also conceptually, it is useful to know how many of these multi--zeta values are linearly independent for fixed transcendentality. The rich structure that emerges has proved fruitful in analyzing amplitudes at tree level.

This question gets more involved at one loop in string theory. As discussed before, the coefficients of the various terms involve graphs of varying modular weights that depend on the string theory that is being considered. While the maximum number of vertices that can arise in a graph is determined by the number of vertex operators, the number of links in these graphs increase as we consider terms at higher orders in the momentum expansion. Thus very heuristically these seem to be the analogs of the multi--zeta values at one loop, with the number of links playing the role of transcendentality. Thus it is interesting to study relations satisfied by these graphs. In particular, knowing how many of them are independent  for a fixed modular weight is helpful in calculating these amplitudes, apart from being important conceptually. One expects a richer structure than that obtained at tree level, which should generalize to higher loops. 

Relations between graphs some of whom require regularization do not conserve the number of links, as was shown for some simple cases in~\cite{Basu:2017nhs}. Deriving such relations in general is expected to be involved, because of boundary contributions from moduli space. On the other hand, we expect relations only between graphs that do not require regularization to follow simply as they do not require regularization. Hence these relations only involve such graphs and the number of links is conserved. Thus for fixed modular weight and fixed number of links, such graphs form a closed subset of all graphs. Hence it is useful to find such relations, which enables us to find a basis and also simplify calculations. 

Such relations have typically been obtained by finding Poisson equations for the graphs and then manipulating them. We shall, on the other hand, find non--trivial relations among such graphs by performing trivial operations starting from appropriate auxiliary graphs~\cite{Basu:2016xrt}. We consider graphs with six links having modular weight four that are are relevant for the $t_8t_8D^4\mathcal{R}^4$ term as an example, but this can be generalized to various other cases.

\begin{figure}[ht]
\begin{center}
\[
\mbox{\begin{picture}(280,180)(0,0)
\includegraphics[scale=.55]{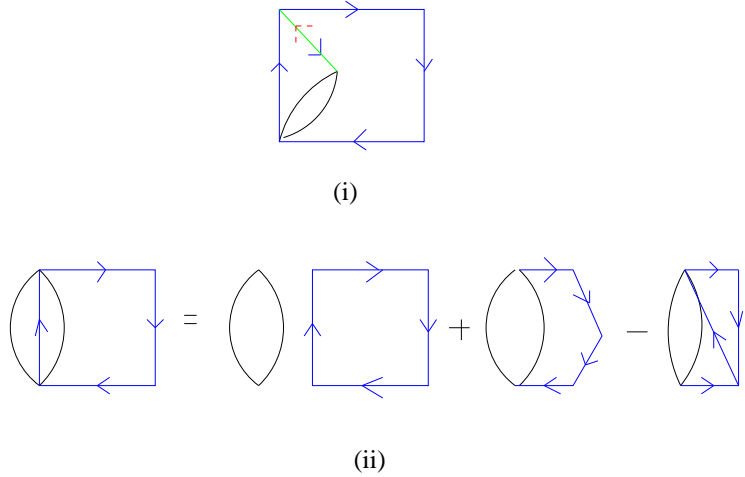}
\end{picture}}
\]
\caption{(i) auxiliary graph, (ii) a relation among graphs}
\end{center}
\end{figure}                

As an example, we start with the auxiliary graph in figure 7. Evaluating it using \C{eigen} for the link involving both $\p G$ and $\bar\p G$ we get a weight four graph with six links, having only $G$ and $\bar\p G$ as the links. On the other hand, evaluating it by moving the $\p G$ to the left and the right and using \C{eigen} we get different graphs. Equating them gives us the relation between graphs in figure 7. Thus we see that completely trivial manipulations using auxiliary graphs yield non--trivial relations between such graphs, where the number of links is conserved.      
We give more relations between such graphs in figure 8, which can be deduced using appropriate auxiliary graphs.

\begin{figure}[ht]
\begin{center}
\[
\mbox{\begin{picture}(280,350)(0,0)
\includegraphics[scale=.65]{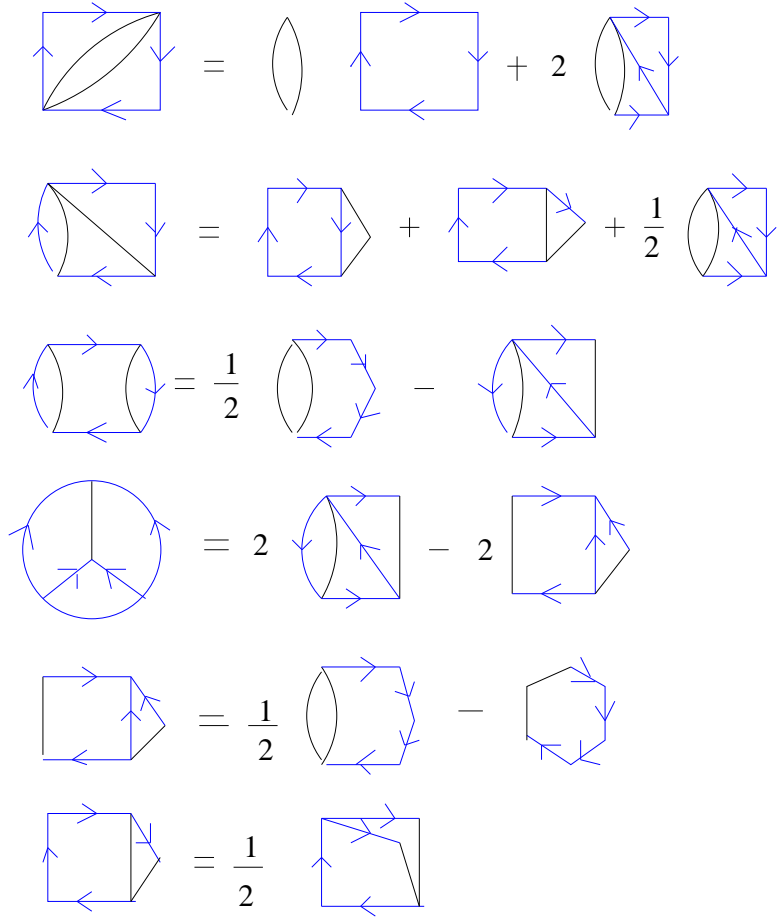}
\end{picture}}
\]
\caption{Some relations among graphs with six links}
\end{center}
\end{figure} 

One can easily deduce many relations involving such graphs which do not require regularization having arbitrary number of links and arbitrary modular weight. An example of such a relation involving weight four graphs with eight links is given in figure 9.   
Thus such relations reduce the number of independent graphs, and are also useful for simplifying the structure of integrands in string amplitudes. A general understanding of relations between various graphs, which do and do not require regularization will be useful, at one loop and beyond.     

\begin{figure}[ht]
\begin{center}
\[
\mbox{\begin{picture}(370,80)(0,0)
\includegraphics[scale=.55]{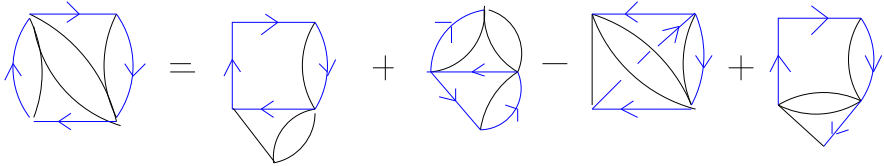}
\end{picture}}
\]
\caption{A relation among graphs with eight links}
\end{center}
\end{figure}

%\bibliographystyle{utphys}
%\bibliography{myrefs}

\providecommand{\href}[2]{#2}\begingroup\raggedright\endgroup

\end{document}